\documentclass[prl,showpacs,preprintnumbers,twocolumn,amsmath,amssymb,superscriptaddress]{revtex4-2}

\usepackage[bookmarks=false]{hyperref}

\usepackage[T1]{fontenc}
\usepackage[applemac]{inputenc}
\usepackage[english]{babel}
\usepackage{graphics}
\usepackage{amssymb}
\usepackage{amsmath}
\usepackage{overpic}
\usepackage{color}

\usepackage[normalem]{ulem}

\renewcommand*{\thefootnote}{\fnsymbol{footnote}}
\newcommand{\be}{\begin{equation}}
\newcommand{\ee}{\end{equation}}
\newcommand{\bea}{\begin{eqnarray}}
\newcommand{\eea}{\end{eqnarray}}

\def\m{\mu}
\def\l{\lambda}

\def\tt{\hat \tau}

\def\s{\sigma}

\def\lb{\label}
\def\pref#1{(\ref{#1})}

\begin{document}

\author{T. Cea}
\email[]{Co-first author.}
\affiliation{IMDEA Nanoscience, C/Faraday 9, 28049 Madrid, Spain}
\author{M. Ruiz-Garc\'ia}
\email[]{Co-first author.}
\affiliation{Department of Physics and Astronomy, University of Pennsylvania, Philadelphia, PA 19104, USA}
\author{L. L. Bonilla}
\affiliation{G. Mill\'an Institute, Fluid Dynamics, Nanoscience and Industrial Mathematics, Universidad Carlos III de Madrid, 28911 Legan\'es, Spain}
\author{F. Guinea}
\affiliation{IMDEA Nanoscience, C/Farady 9, 28049 Madrid, Spain}
\date{\today}

\title{Large-scale critical behavior of the rippling phase transition for graphene membranes }


\begin{abstract}
We analyze the spontaneous rippling of graphene membranes as function of the coupling between lattice deformations and electrons.
 We numerically study a model of an elastic membrane coupled to Dirac fermions. We identify a phase transition from a flat to a rippled configuration of the membrane when increasing the coupling and propose a scaling procedure that allows us to effectively reach arbitrary large system sizes. We find that the critical value of the coupling rapidly decays as the system increases its size, in agreement with the experimental observation of an unavoidable stable rippled state for suspended graphene membranes. This decay turns out to be controlled by a power law with a critical exponent 
 $\sim 1/2$.
\end{abstract}
\maketitle

 \begin{figure}[h!]
\includegraphics[scale=0.3]{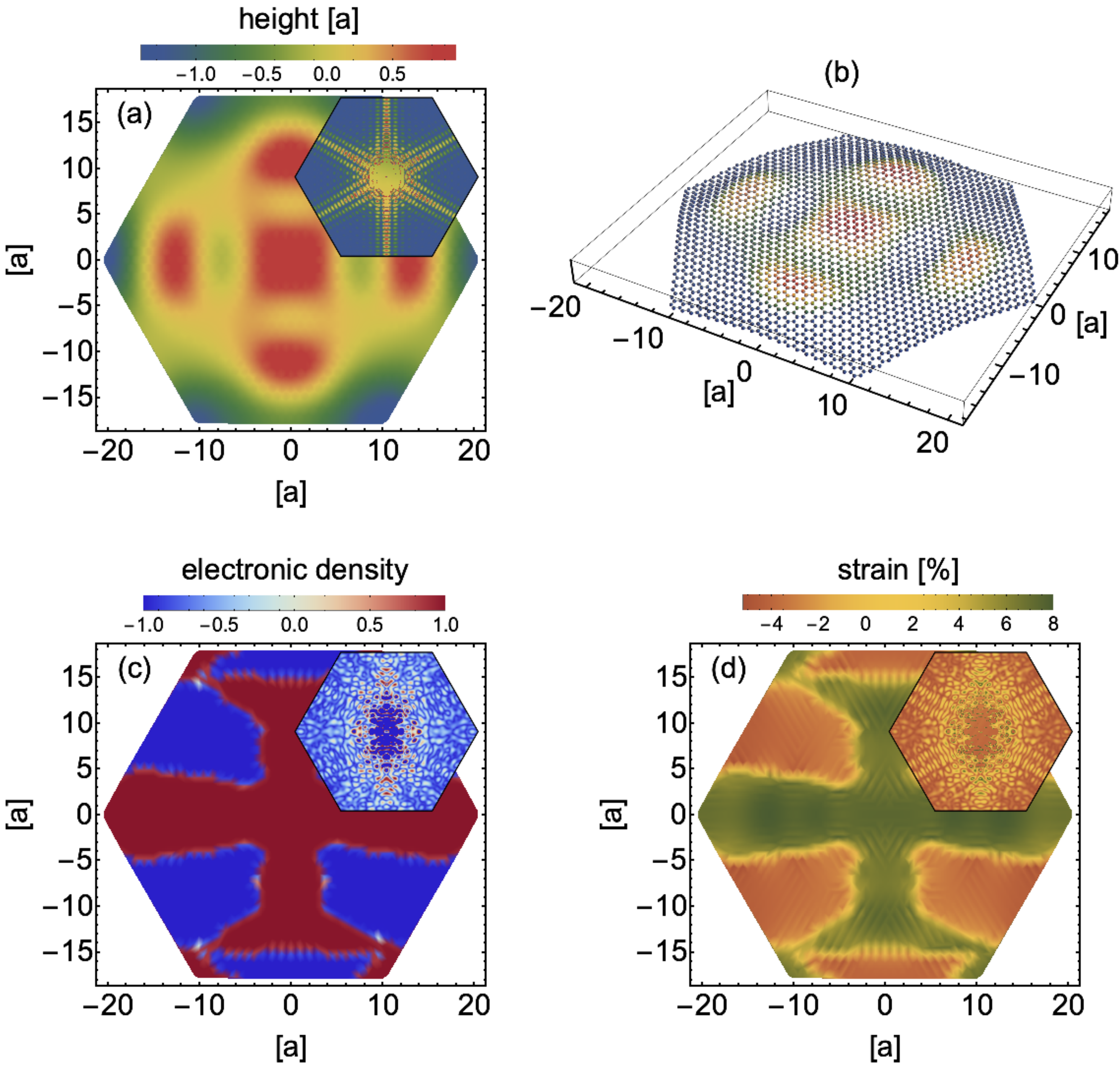}
\caption{
Numerical solution of the elasticity equations coupled to the quantum electronic problem \eqref{spe} for $g=2t_0$. Here we simulate a system of $N_s=2646$ lattice nodes with periodic boundary conditions and lattice constant $a=50a_0$.
Panel (a)-(b): heights distribution shown as a contour plot and 3D plot, respectively.
Panel (c): electronic occupation number counted from half-filling. Panel (d): local strain. The insets show the corresponding Fourier amplitudes in the first Brillouin zone of pristine graphene.
}
\label{fig1}
\end{figure}


\begin{figure}
\includegraphics[scale=0.25]{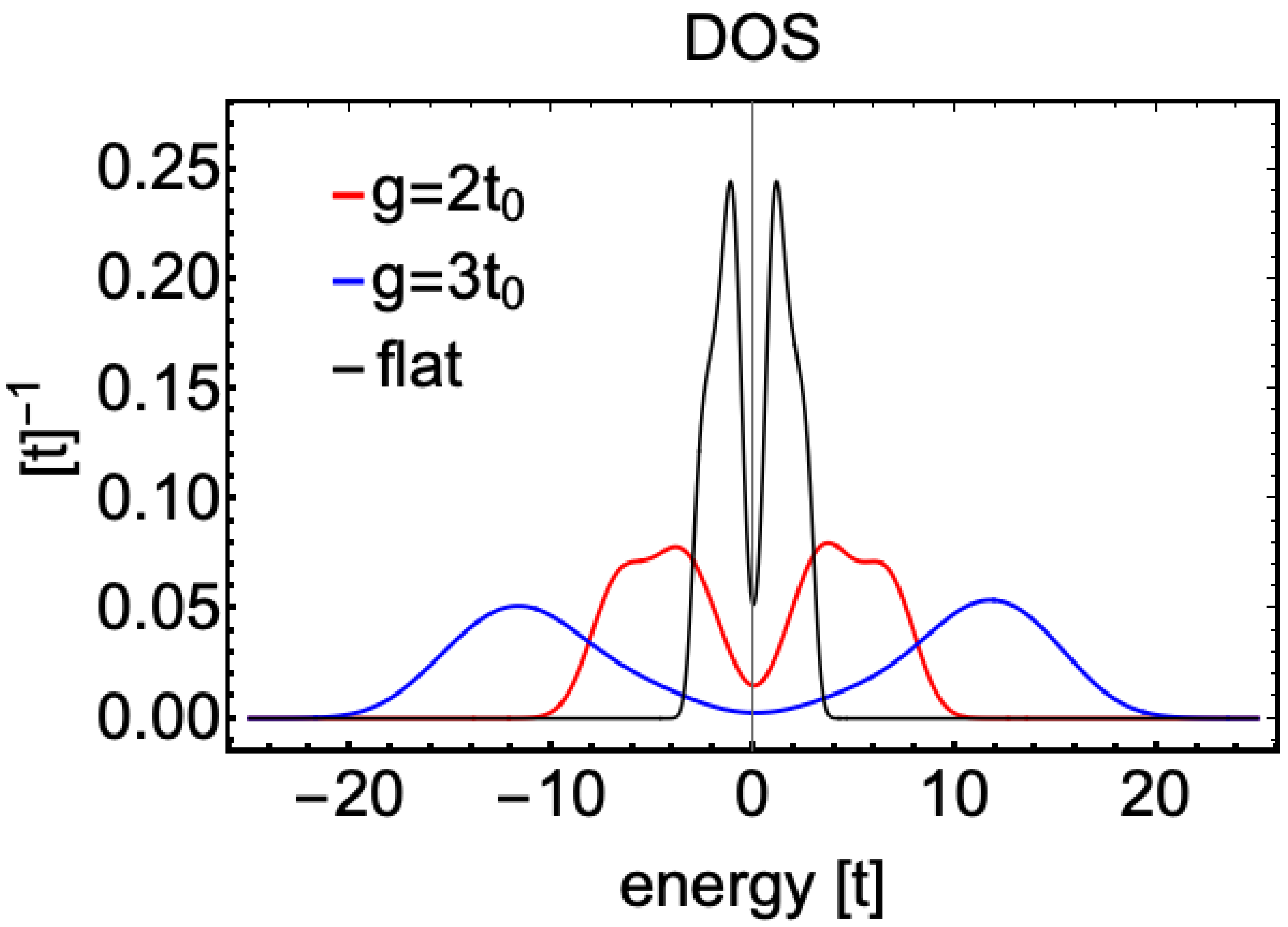}
\caption{Normalized electronic density of the states obtained for different values of the electron-strain coupling $g$. As the coupling increases the rippling is more pronounced, pushing the spectral weight towards higher energies and opening a gap at the Fermi level. The black line, corresponding to the undeformed membrane, represents the typical spectrum of free electrons in graphene in the tight binding approximation. These results correspond to a system of $N_s=2646$ sites with periodic boundary conditions and lattice constant $a=50a_0$, as in Figure \ref{fig1}.
}
\label{fig2}
\end{figure}

 \begin{figure}
\includegraphics[scale=0.35]{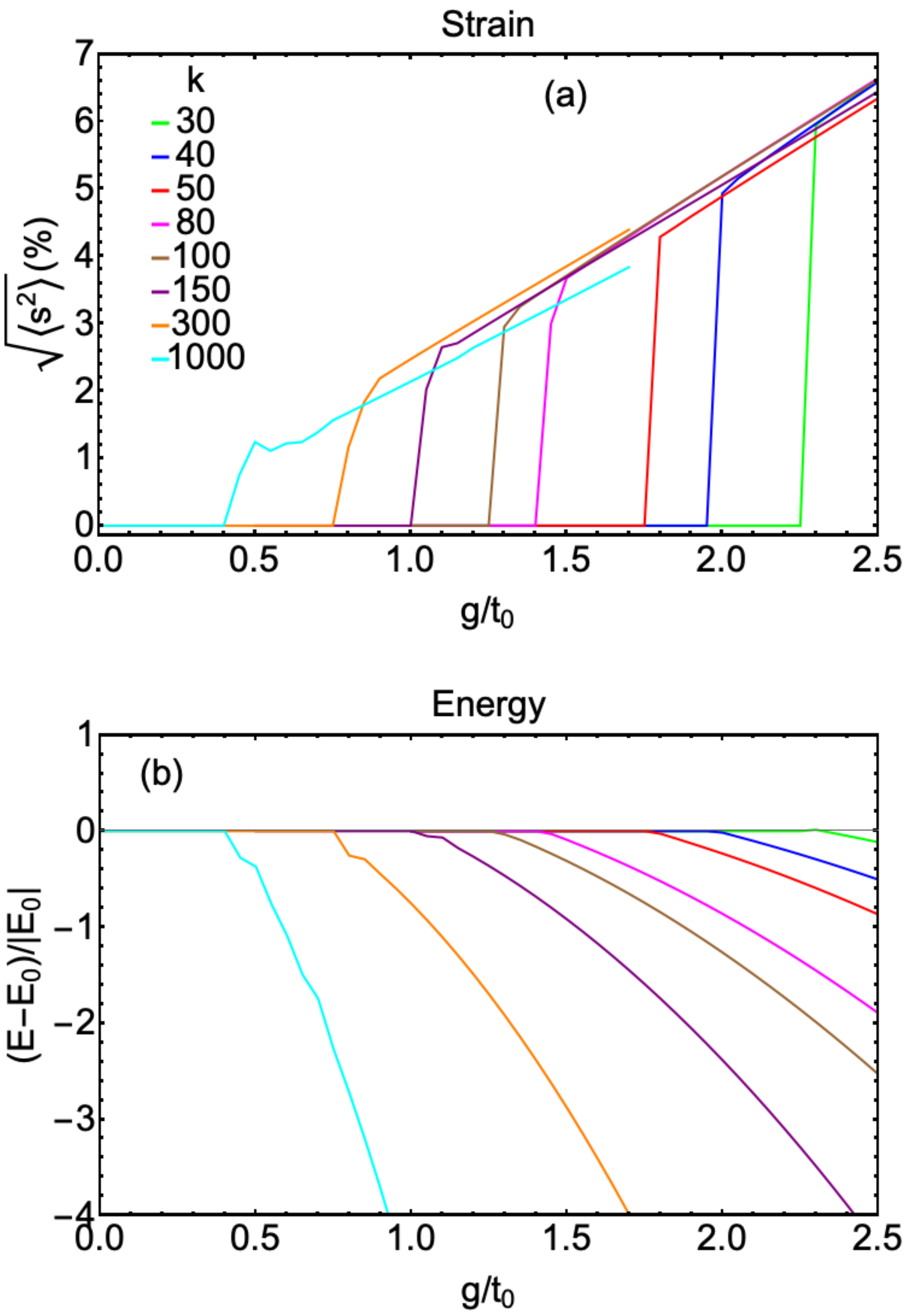}
\caption{Panel (a): geometric average of the local strain
as a function of $g$ for different values of the scaling parameter $k$.
Panel (b): total energy \eqref{Htot} for the same simulations displayed in (a). As $k$ increases the transition occurs for a smaller critical value of $g$.
}
\lb{fig3}
\end{figure}
 \begin{figure}
\includegraphics[scale=0.45]{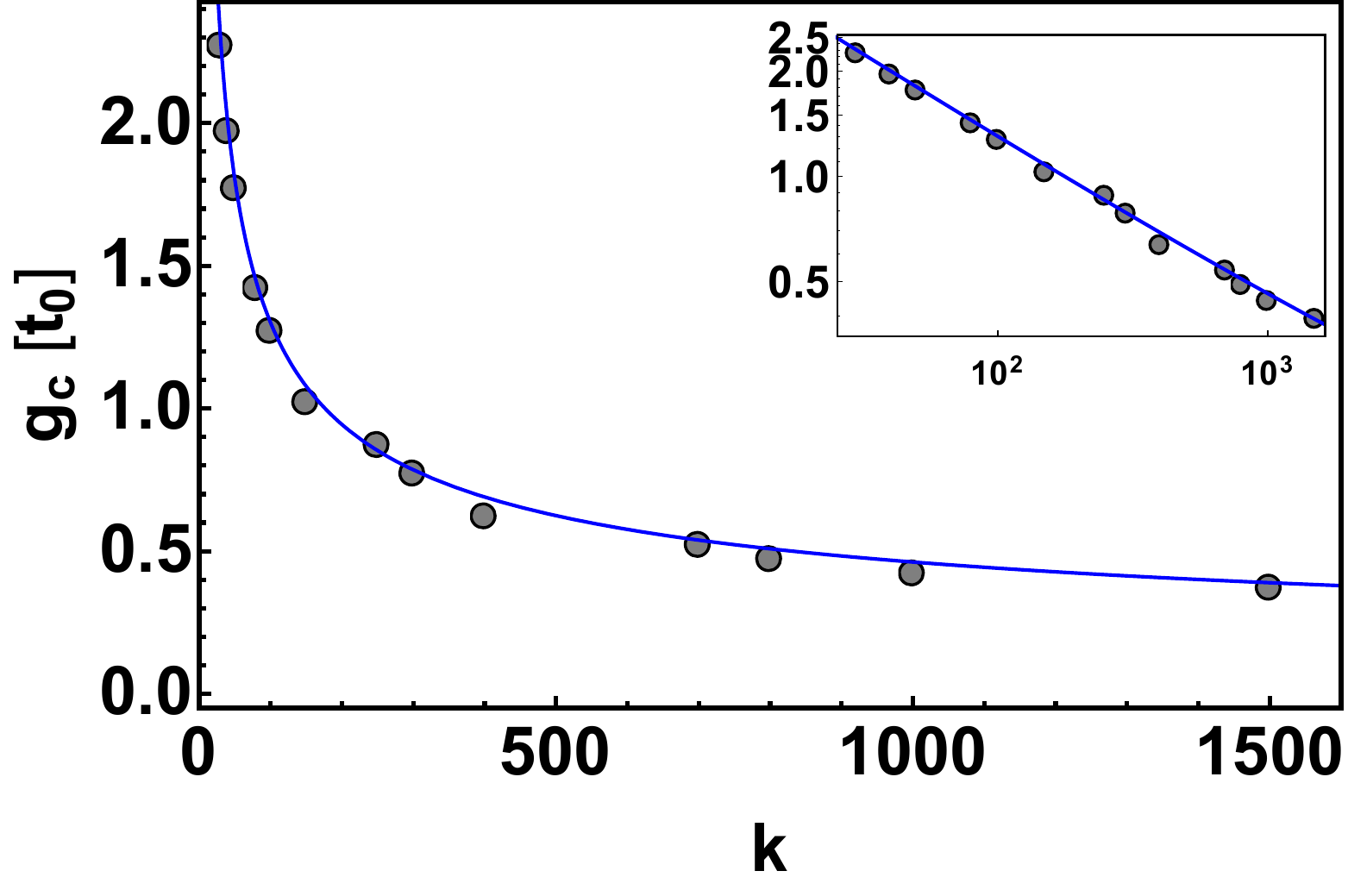}
\caption{
Critical electron-phonon coupling $g_c$ as a function of the scale $k$ (gray dots).
The continuum blue line has been obtained by fitting the data with the power law of Eq.~\eqref{fit}. 
The inset panel represents the data in log-log scale. This result shows how as we consider effectively larger systems  the transition happens at smaller critical values of $g$. Moreover, the dependence on system size displays a power law behavior with a critical exponent of $0.5$.
}
\lb{fig4}
\end{figure}

Graphene isa well studied material ~\cite{Novoselov_science04,Novoselov_nat05,Novoselov_PNAS05,Zhang_nat05}. It has attracted the attention of a broad scientific community due to its unconventional electronic properties~\cite{Novoselov_nat05,castroneto_RevModPhys09,Zhang_nat05,Novoselov_nat05,morozov_prl08,chen_nat08} and its exceptional mechanical properties~\cite{Lee_science08}. In this sense, its atomic thickness makes it the perfect candidate to test the effect that thermal fluctuations can have on its elastic properties~\cite{nelson_book,nelson_JphysF1987,ledoussal_prl92,Letal15,Betal15,LR18,morsheda2019buckling}. However, when suspended, graphene membranes studied using transmission electron microscopy~\cite{meyer_nat07}, show stable ripples, remarkably different from the thermal fluctuations arising on a  flat configuration. Previous works where Ising spins (modelling internal degrees of freedom) are coupled to an elastic membrane show a rich phase diagram \cite{BonillaJSM_2012,ruiz2015ripples,ruiz2016stm}. In $1$D, the model can be analytically solved displaying first and second order phase transitions when controlling the temperature and the interaction between the spins
\cite{ruiz2017bifurcation}.
However, a more realistic approach explaining the origin of rippling in suspended graphene is still undetermined \cite{fasolino_nat07}, and motivates this work.

The out-of-plane deformations of free-standing graphene influence its electronic properties, thereby changing the electrical conductivity~\cite{mariani_prl08,castro_prl10,mariani_prb10} and generating spatially varying gauge potentials~\cite{guinea_prb08,guinea_SSC09,castroneto_RevModPhys09}. These induce charge inhomogeneity~\cite{kim_EPL08} and underlie the formation of electron-hole puddles~\cite{gibertini_prb12}. Previous theoretical studies~\cite{gazit_prb09,sanjose_prl11,guinea_prb14,bonilla2016critical} proposed that the coupling between elastic and electronic degrees of freedom might be at the origin of rippling in graphene, giving rise to a phase transition controlled by the coupling strength. However, in these works, the appearance of ripples is inferred indirectly through a vanishing renormalized bending modulus~\cite{sanjose_prl11} or by a postulated soft mode at finite momentum~\cite{guinea_prb14}. In this work we use a realistic model of suspended graphene membranes. Through numerical simulations we show that stable ripples in suspended graphene membranes can spontaneously arise as a phase transition from a flat state, and that the critical value of the coupling parameter actually tends to zero as the system size increases. This is in agreement with the experimental observation of unavoidable rippling of suspended graphene membranes.

We model a classical elastic membrane on the hexagonal lattice with periodic boundary conditions coupled to a quantum tight binding Hamiltonian for the electrons at zero temperature. The spatial distributions of strain, heights and electronic density are strongly correlated with each other in the rippled phase. Thus, ripples are not triggered by a buckling instability of a clamped or supported finite membrane under tension. Instead, we show that they arise on a large sheet when the electron-strain coupling is large enough to generate stable rippled configurations, characterized by non-homogeneous spatial distributions of strain. We can effectively simulate membranes of arbitrary size by properly defining a scaling parameter $k=a/a_0$, where $a$ is the effective lattice constant and $a_0=2.46${\AA} is that of pristine graphene. In the large system limit the transition happens at very low coupling, which agrees with experimental observations of stable corrugated membranes~\cite{meyer_nat07}. Even more interesting, we find that the critical value of the electron-strain coupling decreases as the power law $\sim k^{-1/2}$ by increasing the scaling size $k$.

The elastic energy of a graphene membrane is given by \cite{landau_book}:
\bea
E_{el}=\frac{1}{2}\int\,d^2\mathbf{r}\left[\kappa\,(\nabla^2 h)^2+\lambda u_{ii}^2+2\mu u_{ij}^2\right], \label{Hel}
\eea
where $\mathbf{u}=(u_x,u_y,h)$ are the in-plane and out-of-plane lattice displacements from the equilibrium positions and $u_{ij}=\frac{1}{2}\left(\partial_iu_j+\partial_ju_i+\partial_ih\partial_jh\right)$ is the strain tensor. $\kappa=0.82$eV is the bending rigidity and $\l,\m$ are the Lam\'e coefficients. Typical values of the Lam\'e coefficients for pristine graphene are: $\lambda_0=19.67$ eV$/a_0^2$ and $\mu_0=57.13$ eV$/a_0^2$. The energy \eqref{Hel} and the displacement vector are discretized on the honeycomb lattice. We assume a direct coupling between the electronic charge density and the strain
and the nearest neighbors tight binding model for the electrons, that can be written as:
\begin{align}
\nonumber
\hat{H}_e&=-t\sum_{\langle \mathbf{R},\mathbf{R'} \rangle \sigma}\left(\hat{a}^{\dagger}_{\mathbf{R}\sigma}\hat{b}_{\mathbf{R'}\sigma}+h.c. \right) \\
&-g\sum_\mathbf{R}(\hat{n}(\mathbf{R}) - n_0)u_{ii}(\mathbf{R}).
\label{H_e}
\end{align}
Here $t$ is the hopping amplitude which, for pristine graphene, is approximatively $t_0=2.7$eV and $\langle \mathbf{R},\mathbf{R'} \rangle$ indicate nearest neighbors sites in the hexagonal lattice.
$\hat{a}$ and $\hat{b}$ are the annihilation operators for a fermion in the $A$ and $B$ sublattices of the honeycomb lattice, respectively,
$\sigma$
is the spin and $g$ is the coupling strength. Recent estimates of $g$ lie in the range:  $g\sim4-50$ eV~\cite{ono_jpsj1966,suzuura_prb02,Petal04,Betal08,Cetal08,choi_prb10,EK10,Petal14}. Finally, the electron density distribution is calculated using $n_0$, the equilibrium occupation number, and $\hat{n}(\mathbf{R})$, the local occupation number, taking the form:
\bea
\hat{n}(\mathbf{R})=
\begin{cases}
\sum_\s \hat{a}^{\dagger}_{\mathbf{R}\sigma}\hat{a}_{\mathbf{R}\sigma}\text{, if $\mathbf{R}\in$ $A$} \\
\sum_\s \hat{b}^{\dagger}_{\mathbf{R}\sigma}\hat{b}_{\mathbf{R}\sigma}\text{, if $\mathbf{R}\in$ $B$}
\end{cases}
\eea
Within this approximation, the graphene Fermi velocity is $v_F= \sqrt{3} t_0 a_0/2\hbar$. Finally, the complete Hamiltonian for the graphene membrane coupled to the electron density distribution is given by:
\bea\lb{Htot}
\hat{H}=E_{el}+\hat{H}_e.
\eea
In the $T=0$ limit, the equilibrium configuration of the membrane minimizes the energy. Since we are treating the electrons in the tight binding approximation, we minimize $\left\langle H \right\rangle$, with respect to both the elastic and electronic degrees of freedom.
We study the Hamiltonian \pref{Htot} within the Born-Oppenheimer (adiabatic) approximation. We treat the displacements as classical fields, whereas we consider the quantum problem for the electrons. To solve the minimization problem, we consider its Euler-Lagrange equations. At zero temperature, the variation of the average energy with respect to the displacement fields yields:
\bea
\frac{\delta E_{el}}{\delta \mathbf{u}(\mathbf{r})}+\left\langle \frac{\delta \hat{H}_{e-ph}}{\delta \mathbf{u}(\mathbf{r})}  \right\rangle=0,
\label{spe}
\eea
where we have used the Hellmann-Feynman theorem to exchange the order in which the quantum average and the functional derivative are computed: $\frac{\delta \left\langle \hat{H} \right\rangle }{\delta \mathbf{u}(\mathbf{r})}=  \left\langle \frac{\delta \hat{H}}{\delta \mathbf{u}(\mathbf{r})}  \right\rangle$.
The procedure followed to solve Eq. \pref{spe} is detailed in Ref. \cite{PRB}. There, the equations are discretized on an hexagonal lattice with $N$ atoms, on the line of the Ref. \cite{carpio_prb2012}. The final equations can be rewritten as the F\"{o}ppl-von K\'arm\'an equations (elasticity of the membrane) plus source terms stemming from the electronic distribution.

We compute the quantum average by diagonalizing the electronic part of the Hamiltonian, with the displacements $\mathbf{u}$ acting as external parameters.  The contribution to the energy is the sum of the lowest half of the eigenvalues. This corresponds to the half filled system, $n_0=1$, with no doping. Then, the corresponding electronic distribution enters the elastic equations as source field.
We solve this nonlinear eigenvalue problem numerically, by starting from an initial configuration of the membrane, $\mathbf{u}_0$, and solving the Eqs.~\eqref{spe} iteratively until reaching the convergence, see details is \cite{PRB}. For $g=0$, the only solution is given by the flat membrane, i.e. zero displacement vectors. For $g\ne0$ the flat configuration is still a solution, but other non flat solutions, $h\ne0$, may appear. We are interested in the possible bifurcations from the flat state as $g$ increases. They correspond to transitions to rippling states.

The main limitation of our approach is the problematic scaling to membranes of realistic sizes. Even though small samples of the order of $10^3$ carbon atoms already give interesting results, see \cite{PRB}, our goal is to understand the behavior of the transition for much bigger membranes. In particular, we aim to study the behavior of the transition for arbitrary large scales, where $g_c$ is the critical value of the coupling constant at which the transition occurs. To circumvent the limitations of simulating a large amount of atoms, we solve Eqs.~\eqref{spe} varying the scaling parameter $k=a/a_0$ (introduced above), which controls the effective size of the sample. Varying $k$ is equivalent to te definition of a new honeycomb lattice where each point does not correspond to a graphene unit cell, but to a coarse grained sample of the original membrane. We scale the terms of our equations so that the elastic energy is independent of the scaling. The Lam\'e coefficients must scale as: $\lambda=\lambda_0/k^2$ and $\mu=\mu_0/k^2$, while the bending energy, $\kappa$, remains constant. Moreover, requiring that the Fermi velocity does not vary upon scaling, we impose that the hopping parameter scales as: $t=t_0/k$.

Figure \ref{fig1} shows a solution of the iterative scheme for a membrane with $N_s=2646$ lattice nodes, periodic boundary conditions, lattice constant $a=50a_0$ and  $g=2t_0$. We include different plots showing the space distribution of the deformation fields (a)-(b) and the charge distribution (c), along with their Fourier transforms (insets). The values of the strain, shown in panel (d), range approximatively from $-5\%$ to $8\%$, in agreement with the experimental observations, and well below the threshold for fracture \cite{meyer_nat07,fasolino_nat07}. Note the correlation of the strain and the charge distribution due to the coupling between the two. To visualize the effect of the electron-strain coupling on the electronic spectrum, in Figure \ref{fig2} we show the electronic density of states for $g/t_0=2$ and $3$, normalized to $1$. The continuum black line represents the Dirac limit corresponding to the flat membrane. As $g$ increases, the spectral weight is pushed to higher energies, up to $\sim 20 t$, while the band edge for free electrons in graphene is $3t$. Furthermore, the data suggest the opening of a gap in the electronic spectrum in the rippled phase, which may be linked to the Anderson localization induced by disorder \cite{anderson_58}.

Figures \ref{fig3} and \ref{fig4} show the rippling transition for systems of different effective sizes, controlled by the scaling parameter $k$. For any value of $k$, as $g$ increases from zero there is a critical value, $g_c$, at which the flat solution loses its stability and the membrane displays rippling. As $k$ increases, this critical value decreases. When plotted against $k$, see figure \ref{fig4}, $g_c$ follows a power law with a critical exponent $\sim 0.5$. A simple fit using the model function:
\begin{equation}\label{fit}
g_c(k)=g_{c,\infty}+C\, k^{-\alpha}, 
\end{equation}
gives: $g_{c,\infty}=0.06t_0$, $C=11.8t_0$ and $\alpha=0.49$.
The small ratio $g_{c,\infty}/t_0$ makes ripples stable even for very small couplings, in agreement with the experimental observations\cite{meyer_nat07} of unavoidable rippling in suspended graphene monolayers. Even when the order parameter seems to jump at the transition, see figure \ref{fig3} (a), these jumps get reduced as we increase $k$ and therefore we think that they may be just a finite size effect. The fact that we find a critical exponent of $\sim 1/2$ when looking at the evolution of $g_c$ versus $k$ suggests that the rippling transition of graphene membranes is of second order. However, we do not have yet a solid theoretical basis to irrefutably establish the order of the transition, an issue which lies outside  the scope of this work.\\\\

The phase transition analyzed above can be cast as a Ginzburg-Landau expansion, using as a set of order parameters amplitudes and wavevectors which describe the lattice distortions in the disordered phase, $\vec{u} ( \vec{r} ) = \sum_{\vec{G}} \vec{u}_{\vec{G}} e^{i \vec{G} \vec{r}}$. The free energy can be written as
\begin{align}
{\cal F} ( &\{ \vec{u}_{\vec{G}} \} ) = \sum_{n=1}^{n = \infty} \frac{c_n D^n}{v_F^{ n- 1 } \tilde{G}_n^{n-3}} \prod_{i = 1, \cdots , n } \vec{G_i } \vec{u}_{\vec{G_i }} \delta \left( \sum_i \vec{G}_i = 0 \right) + 
\nonumber \\
&+\sum_{\vec{G}} \left[ \frac{\lambda}{2} \left( \vec{G} \vec{u}_{\vec{G}} \right)^2 + \right. \nonumber \\
&\left. + \mu \sum_{k=x,y;l=x,y} \frac{\left( G_k  u_{\vec{G} , l} + G_l u_{\vec{G} , k} \right)^2}{2} \right]
\label{GL}
\end{align}
where the parameters $\{ c_n \}$  dimensionless constants of order unity, and $\tilde{G}_n \sim {\rm min}_{i =1 , \cdots , n} ( \{ | \vec{G}_i | \} )$. The term $n=1$ in eq.(\ref{GL}) can be included as a renormalization of the elastic constants. The term $n=2$ is the second order contribution to the total energy in a perturbation expansion on the deformation potential, $D$.  Its value is always negative, and it leads to an inhomogeneous phase for sufficiently large values of the deformation potential. As function of the wavevector,  $\vec{G}$,  this term grows as $| \vec{G} |^3$, while the elastic term goes as $| \vec{G} |^2$. 
Hence, high values of $| \vec{G} |$ are favored. The possible values of $| \vec{G} |$ are limited by the inverse of the lattice constant, so that the inhomogeneous phase will contain deformations at all length scales, up to the lattice constant. This reflects the fact that, in graphene, the electron-phonon coupling is an irrelevant perturbation, in the renormalization group sense, at large length scales. From the numerical point of view, it is quite hard to prove that deformations having small wavelengths are favored, due to the finite size of our systems. However, we checked that the typical size of the ripples decreases by increasing the value of $g$.\\
The long range Coulomb interaction between electrons will modify the transition considered here. The Coulomb interaction suppresses charge accumulation at large length scales, and it is a marginal interaction when compared to the electronic kinetic energy. We do not expect that the Coulomb interaction will change qualitatively the results reported here, as they indicate an instability at small length scales.


In conclusion, we have studied an elastic membrane of graphene coupled to Dirac fermions. We iteratively solved the elasticity equations, discretized on the honeycomb lattice, where the electronic density acts as source field. We find a critical value of the parameter controlling the coupling between deformations and electronic charge, $g_c$, above which a stable ripple phase appears. Upon scaling our equations to account for bigger system sizes, $g_c$ decreases as a power law with critical exponent $\sim 1/2$, until it converges to a fixed point, suggesting that the transition is actually of second order.
  
Our results provide numerical support to previous theoretical works~\cite{gazit_prb09,sanjose_prl11,guinea_prb14,bonilla2016critical}
that proposed the coupling between the electrons and the
elastic degrees of freedom of the membrane
as the driving mechanism at the origin of rippling in suspended graphene sheets. We hope this new perspective and the discovery of the critical exponent can motivate new experimental work shedding light on the physics of rippling in suspended two-dimensional materials.

 \section{Acknowledgements}

M.R-G aknowledges support from the National Science Foundation (USA) and
the Simons Foundation via awards 327939 and 454945. This work has been supported by the FEDER/Ministerio de Ciencia, Innovaci\'on y Universidades -- Agencia Estatal de Investigaci\'on grant MTM2017-84446-C2-2-R (LLB).

\bibliography{Literature}



\end{document}